\documentclass[12pt,A4paper]{article}

\usepackage{epsfig,amssymb}
\usepackage{latexsym}

\begin{document}
\author{A. Guarino$^{{\sf *}}$, R. Scorretti$^{{\sf *}}$, S. Ciliberto$^{{\sf *}}$
and A. Garcimart\'{\i}n$^{{\sf \ddagger }}$  \\
 $^{{\sf *}}${\small
Ecole Normale Sup\'{e}rieure de Lyon, 46 all\'{e}e}\\ d'Italie,
69364 Lyon, France\\ $^{{\sf \ddagger }}${\small Departamento de
F\'{\i}sica, Facultad de}\\ Ciencias, Universidad de Navarra,\\
{\small \ E-31080 Pamplona, Spain.}}
\title{The critical exponents of fracture precursors }
\date{}
\maketitle

\begin{abstract}
The acoustic emission of fracture precursors is measured in
heterogeneous materials. The statistical behaviour of these
precursors is studied as a function of the load features and the
geometry. We find that the time
interval $\delta t$ between events (precursors) and events energies $%
\varepsilon $ are power law distributed and that the exponents of these
power laws depend on the load history and on the material. In contrast, the
cumulated acoustic energy $E$ presents a critical divergency near the
breaking time $\tau $ which is $E\sim \left( \frac{\tau -t}\tau \right)
^{-\gamma }$. The positive exponent $\gamma $ is independent, within error
bars, by all the experimental parameters.
\end{abstract}

\bigskip


\newpage

 Heterogeneous materials are widely studied not only for their
large utility in applications but also because they could give
more insight to our understanding of the role of macroscopic
disorder on material properties. The statistical analysis of the
failure of these materials  is an actual and fundamental problem
which has received a lot of attention in the last decade both
theoretically \cite{Kann}-\cite{golubovic} and experimentally
\cite{stead}-\cite{articolo}.
 When an heterogeneous material is
stretched its evolution toward breaking is characterized by the
appearance of microcracks before their final break-up. Each
microcrack produces an elastic wave which is detectable by a
piezoelectric microphone. The microcraks constitute the so called
precursors of fracture.

The purpose of this letter is just to
describe a detailed statistical analysis of fracture precursors
performed under many different experimental conditions and in
several heterogeneous materials. Analysis of this kind can give
very useful informations for  constructing realistic statistical
models of material failure.
 In our experiments we apply a
pressure P to an heterogeneous sample until failure. The
parameters that we consider are the
elapsed time $\delta t$ between two consecutive events, the acoustic energy $%
\varepsilon $ released by a single microcrack and the acoustic energy
cumulated since the beginning of the loading $E$. In this paper we discuss
the statistical behavior of these parameters as a function of the load
applied to the sample, the material elastic properties and the geometry.

In previous experiments it has been shown \cite{prl,articolo} that if a
quasi-static \footnote{%
A load is considered quasi-static if the load rate is lower than the
relaxation time of the system.} constant pressure rate is imposed, that is $%
P=A_pt$, the sample breaks in a brittle way. In this case the cumulated
acoustic energy $E(t)$ ( i.e. the total energy released up to a time $t$ by
the microfractures) scales with the reduced time or pressure\footnote{%
In this case $P$ and $t$ are proportional.} in the following way:

$$ E\sim ( \frac{\tau -t}{\tau} ) ^{-\gamma } \  \ , $$

where $t$ is time, the critical time $\tau $ is the time at which
the sample breaks and $\gamma =0.27\pm 0.05$ for all the materials
we have checked. In contrast, if a constant strain rate $u=Bt$ is
imposed, a plastic fracture is observed and the released energy
shows no critical behavior. We have also shown that in the case of
a  constant P imposed  to the sample (creep test), the total
energy $E$ becomes, near failure, a function of $t$ and scales as
$E\sim ( \frac{\tau -t}\tau ) ^{-\gamma _c}$. Notably, the
exponent found when a constant stress is applied is the same than
the one corresponding to the case of constant stress rate
\cite{nature}: $\gamma =\gamma _c$. In all of the processes, at
constant pressure and at constant pressure rate, the actual
control parameter for failure seems to be the time. The appearance
of a microcrack seems to be due to a nucleation process
\cite{pomeau,golubovic}, and the probability of nucleation
determines the lifetime $\tau $ of the entire sample. In fact, we
find that $\tau $ is given by the equation:

 $$ \int_0^\tau \frac
1{\tau _o}e^{-(\frac{P_o}P)^4}dt=1 \ \ , $$

 where $P$ is the
pressure and $\tau _o$ and $P_o$ are constants, which depends on
the material and on the geometry \cite{nature}. \newline

In the case of constant load rate ($P=A_pt$ or $u=Bt$) the system has not a
characteristic scale of energy or time: the histogram $N(\varepsilon )$ of
the released energy and the histogram $N(\delta t)$ of the elapsed time $%
\delta t$ between two consecutive events reveal power laws, i.e. $%
N(\varepsilon )\thicksim \varepsilon ^{-\beta }$ and $N(\delta
t)\thicksim \delta t^{-\alpha }$. The exponents $\alpha $, $\beta
$ and $\gamma $ do not depend on the load rate $A_p$ or $B$
\cite{prl,articolo}. In this paper we are interested in studying
the exponents  in different geometries and when a constant (creep
test), cyclic or erratic load are imposed.\newline

The tests are performed by monitoring the acoustic emission (AE)
released before the final break-up of a sample on a high pressure
chamber (HPC) machine. The sample separates two chambers and a
pressure difference $P$ is imposed between them. A sketch of the
apparatus is shown in fig. 1a and 1b. We have prepared circular
wood (Young modulus$ Y=2 \ 10^8 N/m^2$)  and fiberglass samples of
22 cm diameter and 4 mm thickness. The Young modulus  of these
materials are $Y=2 \ 10^8 N/m^2$ for wood and $ Y=2 \ 10^8 N/m^2$
for fiberglass.
 The AE
consists of ultrasound bursts (events) produced by the formation
of microcracks inside the sample. For each AE event, we record
the energy $\varepsilon $ detected by the four microphones \footnote{%
The energy is defined as the integral of the sum of the squared signals.},
the place where it was originated, the time at which the event was detected
and the instantaneous pressure and displacement at the center of the sample.
We are able to record up to 33 events per second. The experimental apparatus
is the same that has been used to obtain the previously cited results; a
more detailed description of the experimental methods can be found in \cite
{prl,articolo}.\newline

To check the dependence of $\alpha $, $\beta $ and $\gamma $ on the
geometry, we used a classical tensile machine, fig 1c. The force applied to
the sample is slowly and constantly increased up to the final break-down of
the sample. During the load we measure the applied force $F$, the strain,
the AE produced by microcracks and the time at which the event was detected.
The samples have a rectangular shape, with a length of 29 cm, a height of 20
cm and a thickness of 4 mm. More details of the experimental setup can be
found in \cite{jf}.\newline

In the experiments performed with the first apparatus (HPC), power
laws are obtained for the distributions of $\epsilon $ and of
$\delta t$. As an example of two typical distributions obtained at
constant imposed pressure, we plot in  fig 2a) and 2b) $N(\delta
t)$ and $N(\delta \epsilon)$ respectively. The exponents of these
power laws ($\alpha _c$ for energies and $\beta _c$ for times)
depend on $P$. In fig. 2c, $\alpha _c$ and $\beta _c$ are plotted
versus $P$. Note that both exponents grow with pressure. We
observe that the rate of emissions increases with pressure, so
that the weight of big values of $\delta t$ decreases. This
explains the fact that $\beta _c$ grows with pressure. We have
compared the histograms of energy $\epsilon $ for several
pressures, and we noticed that the number of high-energy emissions
is almost the same, while the number of low-energy emissions
increase with pressure, so that the exponent $\alpha _c$ increases
as well. Moreover, as the pressure increases, the exponents
$\alpha _c$ and $\beta _c$ attain the values $\alpha =1.9\pm 0.1$
and $\beta =1.51\pm 0.05$ obtained in the case of a constant
loading rate\cite{articolo}. We imposed to the sample a cyclic and
an erratic load, which are plotted as a function of time in figure
3a and 3b respectively. Power laws are obtained for the
distributions of $\epsilon $ and for $\delta t$. The exponents of
these power laws do not depend on the load behavior; their value
is the same of that at constant loading rate. These and previous
results \cite{prl,articolo}, allows us to state that if
$\frac{dP}{dt}\neq 0$, the
histograms of the released energy $\varepsilon$ and of the time intervals $%
\delta t$ do not depend on the load history. The fact that $\alpha $ and $%
\beta $ do not depend on $\frac{dP}{dt}$ seems to be in contrast
with the fact that $\alpha _c$ and $\beta _c$ depend on $P$. This
result can be interpreted by considering that the microcracks
formation process is not the same when $\frac{dP}{dt}=0$ and
$\frac{dP}{dt}\neq 0$. In the former case, imposed constant $P$,
the mechanism of microcrack nucleation is the dominant
one and the nucleation time depend on pressure. In the other case, $\frac{dP%
}{dt}\neq 0$, the dominant mechanism is not the nucleation but the fact
that, when pressure increases as a function of time, several parts of the
sample may have to support a pressure larger than the local critical stress
to break bonds. The fact that at high constant pressure $\alpha _c$ and $%
\beta _c$ recover the value $\alpha _c$ and $\beta _c$ has a simple
explanation. Indeed, in order to reach a very high pressure $P_h$, $\frac{dP%
}{dt}$ is different from zero for a time interval which is
comparable or even larger than the time interval spent at constant
pressure $P_h$. Thus at high constant pressure the system is close
to the case $\frac{dP}{dt}\neq 0$.
\newline

Using the cyclic and the erratic pressure, plotted respectively in
fig. 3a and 3b, we can check the dependence of $\gamma $ on the
history of the
sample, i.e. on the behavior of the imposed pressure. The cumulated energy $%
E $ for the cyclic and the erratic pressure, shown in fig 3a and
3b as a function of $t$, is plotted in log-log scale as a function
of the reduced parameter $\frac{\tau -t}\tau $ in fig 4a and 4b
respectively. We observe that, in spite of  the fluctuations due
to the strong oscillations of the applied pressure, near the final
break-up the energy $E$, as a function of $\frac{\tau -t}\tau $,
is fitted by a power law with $\gamma\simeq 0.27 \pm 0.02$.
 In
fig.4c),  $E$ measured  when a constant pressure is applied to the
sample is plotted  as a function of $\frac{\tau -t}\tau $. A power
law is found in this case too \cite{nature}.  The exponent $\gamma
$ is, within error bars, the same in the three cases. Hence it
depends neither on the applied pressure history nor on the
material
\cite{prl,articolo,nature}.%
\newline

Further, experiments made with the tensile machine show that
$\gamma $ is independent on the geometry. In fact we observe that
the behavior of the energy near the fracture as a function of
$\left( \frac{\tau -t}\tau \right) $ is still a power law of
exponent $\gamma\simeq 0.27 $, as shown in figure 4d.

\smallskip

Considering the experimental data here presented and those already published
\cite{prl,articolo}, we claim that, if a load is imposed to an heterogeneous
material, power laws are obtained for the histograms of the released energy $%
\varepsilon $ and of the time intervals $\delta t$. The exponents
of these power laws depend on the material and, if
$\frac{dP}{dt}=0$ , on the applied pressure $P$. In contrast, at
imposed pressure, the behavior of the cumulated energy $E$ near
the final breaking point does not depend on the load, on the
geometry and on the material \cite{prl,articolo,nature}. We find
that time is the  control parameter  of the system and that $E\sim
\left( \frac{\tau -t}\tau \right) ^{-\gamma }$, where the critical
exponent is $\gamma =0.27\pm 0.05$. These results are quite
similar to those obtained with numerical simulations on a
democratic bundle fiber model with thermal noise. These facts and
the observed  dependence of $\tau$ on $P$  allow us to conclude
that microcrack nucleation process \cite{pomeau,golubovic} plays a
fundamental role in the entire dynamics of the system. These are
very useful informations in order to construct a realistic
statistical model of heterogenous material failure.

\smallskip

\begin{figure}

\centerline{\epsfysize=0.9\linewidth \epsffile{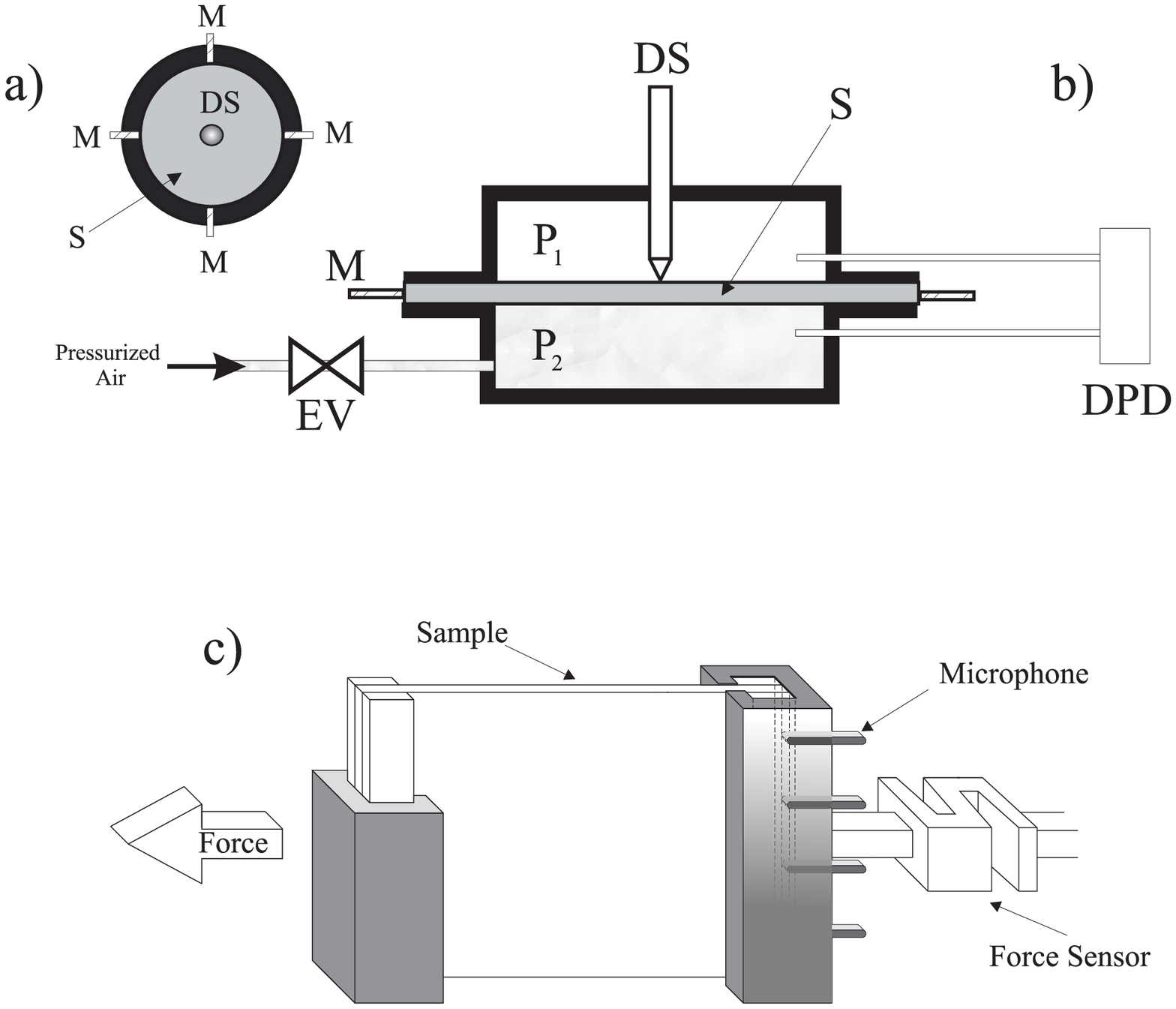}}

 \caption{
 {\bf a, b) } Sketch of the high pressure chamber (HPC)
apparatus. S is the sample, DS is the inductive displacement
sensor (which has a sensitivity of the order of 1 $\mu $m). M are
the four wide-band piezoelectric microphones. P=P$_1$-P$_2$ is the
pressure supported by the sample. P is measured by a differential
pressure sensor ( sensitivity 0.002 atm) that is not represented
here. EV is the electronic valve which controls P via the feedback
control system Ctrl . HPR is the high-pressure air reservoir. {\bf
c)} Sketch of the tensile machine. An uniaxial force, which is
measured by a piezoresistive sensor, is applied to the sample by a
stepping motor. Four wide-band piezoelectric microphones measure
the acoustic emissions emitted by the sample. Experiments have
been done using rectangular (20 x 29 cm) wood samples of 4 mm
thickness. The whole apparatus is surrounded by a Faraday screen.}
\label{fig:exp}
\end{figure}

\begin{figure}
\centerline{\epsfysize=1.1\linewidth \epsffile{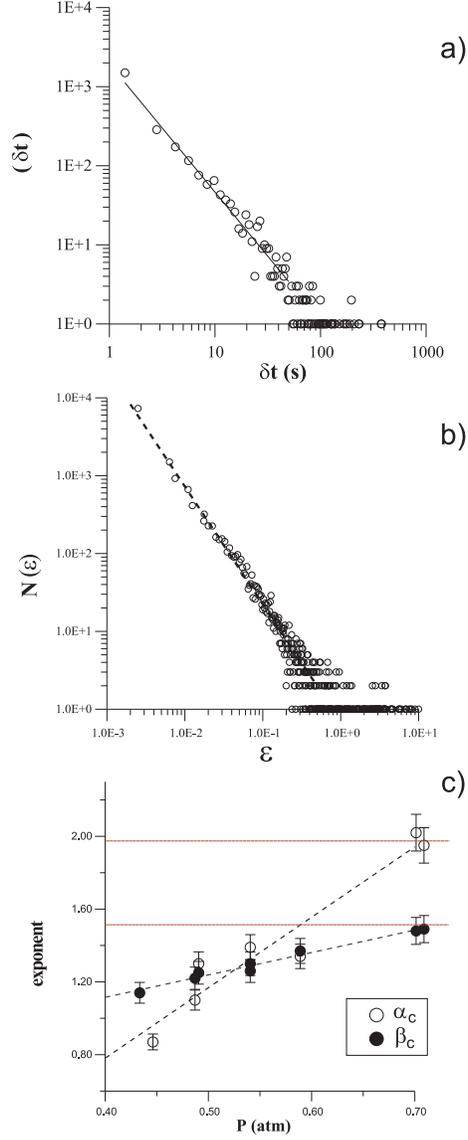}}

\caption{
 {\bf a,b)} Two typical time $\delta t$ and energy
$\varepsilon $
distributions obtained at imposed constant pressure ($P=0.56$ $atm$). {\bf c)%
} The exponents $\alpha _c$ (empty circles) and $\beta _c$ (black
points), plotted as a function of the value of the imposed
constant pressure. Note that as the pressure increases, the values
of the exponents tend to those obtained in the case of constant
pressure rate. The error bars represent the statistical
uncertainty.}
 \label{fig:hist}
\end{figure}

\begin{figure}
\centerline{\epsfysize=0.9\linewidth \epsffile{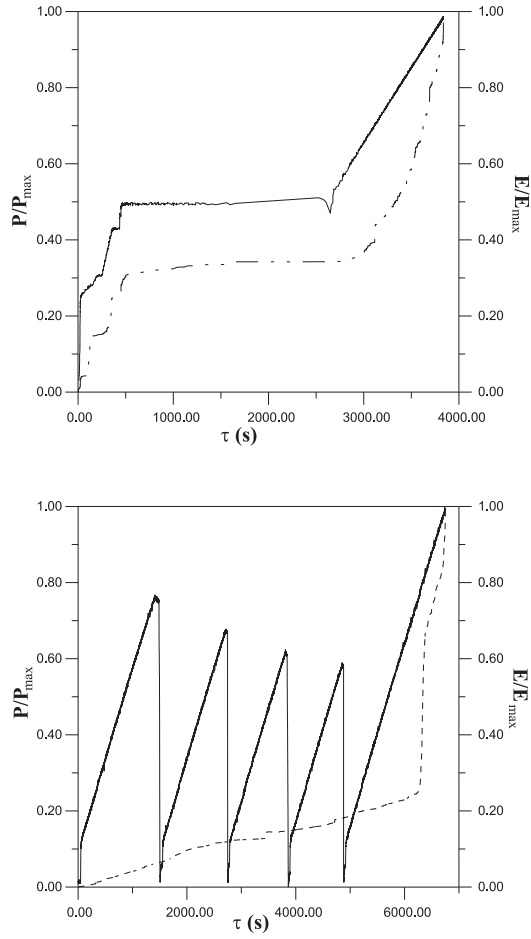}}

\caption{
  The imposed pressure, normalised at P$_{\max }$, and
the cumulated
energy $E$, normalized to E$_{\max }$, are plotted as a function of time $t$%
. a) An example of erratic pressure. b) A cyclic pressure.}
 \label{fig:press}
\end{figure}

\begin{figure}

\centerline{\epsfysize=0.9\linewidth \epsffile{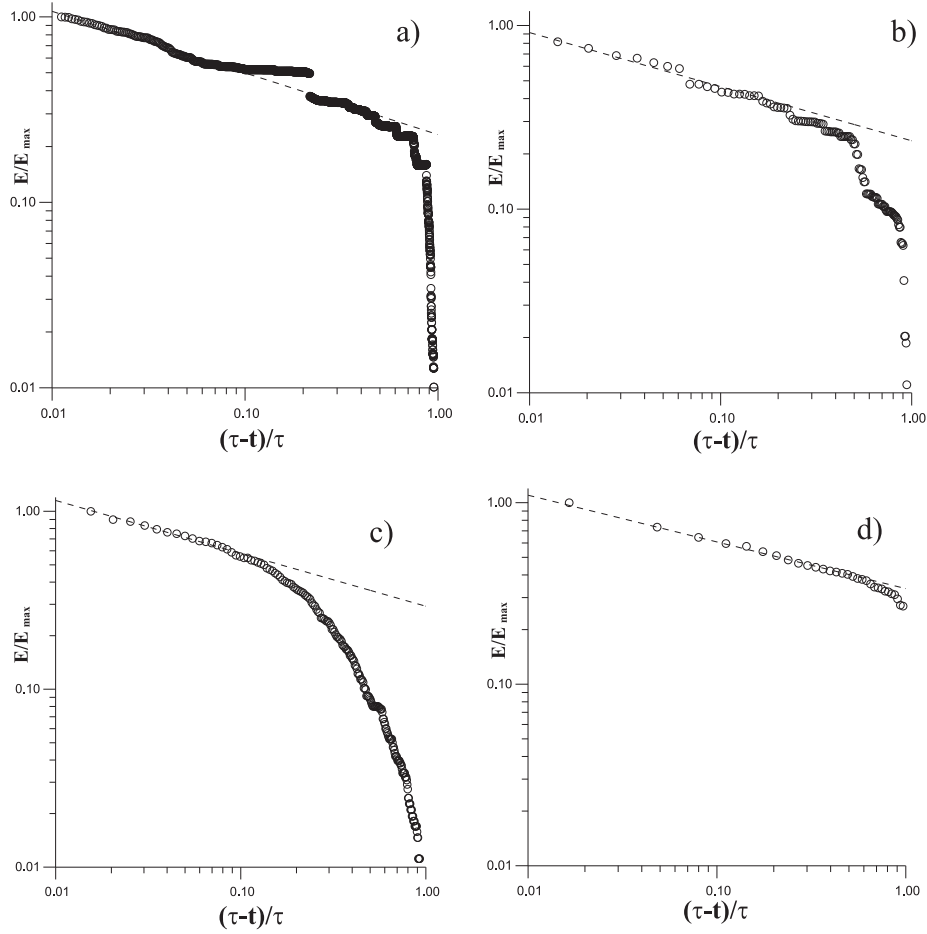}}
\caption{
  The cumulated energy $E$, normalized to $E_{\max }$, as a
function of the reduced control parameter $\frac{\tau -t}\tau $ at
the neighborhood of the fracture point. Figure d) represent the
measure taken, at imposed constant rate force, on the tensile
machine. The other figures represent measures made on the HPC
apparatus at : imposed constant pressure (c), imposed cyclic
pressure (a) and imposed erratic pressure (b). The dotted lines
are the fit $E=E_0\left( \frac{\tau -t}\tau \right) ^{-\gamma }$.
The exponents found are: $\gamma =0.29$ (a), $\gamma =0.25$ (b),
$\gamma =0.29$ (c) and $\gamma =0.27$ (d). In the case of a
constant pressure rate (on the HPC machine) the same law has been
found \cite{articolo,prl}}
 \label{fig:critic}
\end{figure}

\end{document}